\documentclass[12pt]{article}
\usepackage{amsfonts}
\usepackage{amssymb}

\def\hybrid{\topmargin -20pt    \oddsidemargin 0pt
        \headheight 0pt \headsep 0pt
        \textwidth 6.25in       
        \textheight 9.25in       
        \marginparwidth .875in
        \parskip 5pt plus 1pt   \jot = 1.5ex}

\hybrid

\def\baselinestretch{1.2}

\catcode`\@=11

\def\marginnote#1{}
\newcount\hour
\newcount\minute
\newtoks\amorpm
\hour=\time\divide\hour by60
\minute=\time{\multiply\hour by60 \global\advance\minute by-\hour}
\edef\standardtime{{\ifnum\hour<12 \global\amorpm={am}%
        \else\global\amorpm={pm}\advance\hour by-12 \fi
        \ifnum\hour=0 \hour=12 \fi
        \number\hour:\ifnum\minute<10 0\fi\number\minute\the\amorpm}}
\edef\militarytime{\number\hour:\ifnum\minute<10 0\fi\number\minute}

\def\draftlabel#1{{\@bsphack\if@filesw {\let\thepage\relax
   \xdef\@gtempa{\write\@auxout{\string
      \newlabel{#1}{{\@currentlabel}{\thepage}}}}}\@gtempa
   \if@nobreak \ifvmode\nobreak\fi\fi\fi\@esphack}
        \gdef\@eqnlabel{#1}}
\def\@eqnlabel{}
\def\@vacuum{}
\def\draftmarginnote#1{\marginpar{\raggedright\scriptsize\tt#1}}

\def\draft{\oddsidemargin -.5truein
        \def\@oddfoot{\sl preliminary draft \hfil
        \rm\thepage\hfil\sl\today\quad\militarytime}
        \let\@evenfoot\@oddfoot \overfullrule 3pt
        \let\label=\draftlabel
        \let\marginnote=\draftmarginnote
   \def\@eqnnum{(\theequation)\rlap{\kern\marginparsep\tt\@eqnlabel}%
\global\let\@eqnlabel\@vacuum}  }

\def\preprint{\twocolumn\sloppy\flushbottom\parindent 2em
        \leftmargini 2em\leftmarginv .5em\leftmarginvi .5em
        \oddsidemargin -.5in    \evensidemargin -.5in
        \columnsep .4in \footheight 0pt
        \textwidth 10.in        \topmargin  -.4in
        \headheight 12pt \topskip .4in
        \textheight 6.9in \footskip 0pt
        \def\@oddhead{\thepage\hfil\addtocounter{page}{1}\thepage}
        \let\@evenhead\@oddhead \def\@oddfoot{} \def\@evenfoot{} }

\def\numberbysection{\@addtoreset{equation}{section}
        \def\theequation{\thesection.\arabic{equation}}}

\def\underline#1{\relax\ifmmode\@@underline#1\else
        $\@@underline{\hbox{#1}}$\relax\fi}

\def\titlepage{\@restonecolfalse\if@twocolumn\@restonecoltrue\onecolumn
     \else \newpage \fi \thispagestyle{empty}\c@page\z@
        \def\thefootnote{\fnsymbol{footnote}} }

\def\endtitlepage{\if@restonecol\twocolumn \else \newpage \fi
        \def\thefootnote{\arabic{footnote}}
        \setcounter{footnote}{0}}  

\catcode`@=12
\relax

\def\figcap{\section*{Figure Captions\markboth
        {FIGURECAPTIONS}{FIGURECAPTIONS}}\list
        {Figure \arabic{enumi}:\hfill}{\settowidth\labelwidth{Figure
999:}
        \leftmargin\labelwidth
        \advance\leftmargin\labelsep\usecounter{enumi}}}
 \relax
\def\tablecap{\section*{Table Captions\markboth
        {TABLECAPTIONS}{TABLECAPTIONS}}\list
        {Table \arabic{enumi}:\hfill}{\settowidth\labelwidth{Table
999:}
        \leftmargin\labelwidth
        \advance\leftmargin\labelsep\usecounter{enumi}}}
 \relax
\def\reflist{\section*{References\markboth
        {REFLIST}{REFLIST}}\list
        {[\arabic{enumi}]\hfill}{\settowidth\labelwidth{[999]}
        \leftmargin\labelwidth
        \advance\leftmargin\labelsep\usecounter{enumi}}}
 \relax
\makeatletter
\newcounter{pubctr}
\def\publist{\@ifnextchar[{\@publist}{\@@publist}}
\def\@publist[#1]{\list
        {[\arabic{pubctr}]\hfill}{\settowidth\labelwidth{[999]}
        \leftmargin\labelwidth
        \advance\leftmargin\labelsep
        \@nmbrlisttrue\def\@listctr{pubctr}
        \setcounter{pubctr}{#1}\addtocounter{pubctr}{-1}}}
\def\@@publist{\list
        {[\arabic{pubctr}]\hfill}{\settowidth\labelwidth{[999]}
        \leftmargin\labelwidth
        \advance\leftmargin\labelsep
        \@nmbrlisttrue\def\@listctr{pubctr}}}
 \relax
\makeatother
\newskip\humongous \humongous=0pt plus 1000pt minus 1000pt

\newif\ifdtup

\relax

\def\be{\begin{equation}}
\def\ee{\end{equation}}
\def\ba{\begin{eqnarray}}
\def\ea{\end{eqnarray}}

\def\del{\partial}

\def\r{\rho}
\def\a{\alpha}

\def\G{\Gamma}
\def\d{\delta}
\def\D{\Delta}
\def\e{\epsilon}

\def\P{\Pi}

\def\th{\theta}

\def\m{\mu}

\def\om{\omega}

\def\L{\Lambda}
\def\s{\sigma}

\def\cN{{\cal N}}

\def\no{\noindent}

\def\qq{\qquad}
\def\bl{\bigl}
\def\br{\bigr}

\def\IR{\relax{\rm I\kern-.18em R}}

\def \ha {{1\over 2}}

\def \ov {\over}

\def\IR{\relax{\rm I\kern-.18em R}}
\def\inv{^{\raise.15ex\hbox{${\scriptscriptstyle -}$}\kern-.05em 1}}

\begin{document}

\renewcommand{\theequation}{\thesection.\arabic{equation}}

\newcommand{\beq}{\begin{equation}}
\newcommand{\eeq}[1]{\label{#1}\end{equation}}
\newcommand{\ber}{\begin{eqnarray}}
\newcommand{\eer}[1]{\label{#1}\end{eqnarray}}
\newcommand{\eqn}[1]{(\ref{#1})}
\begin{titlepage}
\begin{center}

\hfill hep--th/0305109\\
\hfill May 2003\\

\vskip .8in

{\large \bf The exact description of NS5-branes in the Penrose limit}

\vskip 0.6in

{\bf Konstadinos Sfetsos}

{\em Department of Engineering Sciences, University of Patras\\
26110 Patras, Greece\\
\footnotesize{\tt sfetsos@des.upatras.gr, mail.cern.ch}}\\

\end{center}

\vskip .4in

\centerline{\bf Abstract}

\no
We construct plane wave backgrounds with time-dependent profiles
corresponding to Penrose limits
of NS5-branes with transverse space 
symmetry group broken from $SO(4)$ to $SO(2)\times Z_N$. We
identify the corresponding exact theory 
as the five-dimensional Logarithmic Conformal Field Theory (CFT)
arising from the contraction of the 
$SU(2)_{N'}/U(1)\times SL(2,R)_{-N}$ exact CFT, times $\IR^5$. 
We study several general aspects and construct the free field representation 
in this theory.
String propagation and spectra are also considered 
and explicitly solved in the light-cone gauge.

\noindent

\vskip 1in
\noindent
\end{titlepage}
\vfill
\eject

\def\baselinestretch{1.2}
\baselineskip 17.5 pt
\noindent

\section{Introduction}

The purpose of the present paper is to provide and analyze the first example 
of a Logarithmic CFT arising as the exact description of certain supergravity
brane configurations in the Penrose limit \cite{Pen}
and to also study string propagation in this background.
Novel Logarithmic CFTs arose recently 
in the large level $N$ limit of coset conformal field
theories for compact groups combined with a free time-like boson \cite{BS1}
and provided the exact description of various PP-waves configurations (with
no brane interpretation however).
It was shown quite generally in \cite{BS1} that,
at the level of the chiral algebra involving
the compact parafermions \cite{ZF,claspara}
and the $U(1)$-current corresponding to the 
free time-like boson, the Penrose limiting procedure gives rise to a 
Saletan-type contraction (for mathematical details in the case of Lie groups,
see \cite{Gilmore}) 
and to a Logarithmic CFT theory \cite{Gur}, \cite{BK}
(for earlier work see \cite{gardy};
for a review and more references see \cite{Flrev}). 

This type of Penrose limits and contractions 
have already appeared in string theory some years ago where they were 
used to construct plane wave solutions starting from WZW model
current algebra theories or gauged WZW model coset theories. The 
prototype example is the plane wave solution of \cite{NW},
corresponding to a WZW model for the non-semisimple group $H_4$,
which is the Penrose limit of the background corresponding to the 
$SU(2)_N\times U(1)_{-N}$ current algebra theory \cite{sfe22} (for 
construction of plane waves either directly as WZW and 
gauged WZW models or by contractions see \cite{Olive}-\cite{Antoniadis}).
The interest in plane wave solutions arising in string and M-theory has been
recently revived following the construction of a maximally supersymmetric
plane wave solution of type-IIB supergravity \cite{Blau1}, the fact that, 
it can be obtained as a Penrose limit of the maximally supersymmetric
vacuum solution, $AdS_5 \times S^5$ \cite{Blau2} and that, string
theory is exactly solvable in this plane wave background \cite{metsaev},
which allows to extend the AdS/CFT
correspondence in a non-trivial way
to include the effect of highly massive string states \cite{BMN}.
In accordance with experience and general expectations, the superalgebra
for the $AdS_5 \times S^5$ background contracts to the
corresponding plane wave
superalgebra, as it has been explicitly demonstrated in \cite{superal}.   
These more recent developments are in accordance with the fact that, 
the Penrose limiting procedure can be straightforwardly
generalized to supergravity theories \cite{Gueven}.

The organization of this  paper is as follows: In section 2 we review 
the relevant aspects of certain 
NS5-brane configurations which will serve as the starting point 
in our constructions.
In the supersymmetric case they have the distinct feature that they preserve
an $SO(2)\times Z_N$ subgroup of the $SO(4)$ symmetry group of the transverse
to the branes space $\IR^4$. We also review the non-extremal extension of 
this solution, representing the most general non-extremal rotating 
NS5-brane solution. In section 3 we systematically construct plane wave
backgrounds with time dependent profiles 
by performing several different Penrose limits. In section 4 
we identify the exact CFT corresponding to these plane waves as the contraction
of the $SU(2)_{N'}/U(1) \times SL(2,\IR)_{-N}$ theory.
The result is a one parameter family of 
five-dimensional Logarithmic CFTs.  For a particular member of this family
the theory degenerates into the four-dimensional current algebra for 
the non-semisimple group $H_4$ times a $U(1)$ factor. In section 5 we 
show that it is possible to explicitly solve string theory in 
the light-cone gauge, even though
our plane wave backgrounds have time dependent profiles.
We end our paper in section 6 
with concluding remarks and directions for feature research.


\section{The background}
\setcounter{equation}{0}

The most general solution of either type-II or Heterotic string theory 
representing the gravitational field and flux of $N$ $NS5$-branes is given by 
\cite{hostro}
\ba
&& ds^2 = \sum_{a=1}^5 dy_a^2-dt^2 + H\sum_{i=1}^4 dx_i^2 \ ,
\nonumber\\
&&H_{ijk}=\e_{ijkl}\del_l H \ , 
\label{skh}\\
&& e^{2\Phi} = H \ ,
\nonumber
\ea
where the coordinates $y_a$'s and $t$ parametrize the world-volume 
of the branes and the $x_i$'s the directions transverse to the branes 
space which is $\IR^4$.
The function $H$ is harmonic in $\IR^4$. Demanding that asymptotically the 
space is flat we have in general
\be 
H= 1 + \sum_{i=0}^{N-1} {\a'\ov |{\bf x}-{\bf x}_i|^2}\ ,
\label{hkhhh}
\ee
where ${\bf x}_i$ denote the locations of the branes in $\IR^4$.
Such configurations preserve sixteen supercharges.
A generic choice for the centers breaks the $SO(4)$ symmetry completely.
In this paper we will consider NS5-brane configurations preserving an 
$SO(2)\times Z_N$ subgroup of $SO(4)$. In particular, we will consider
backgrounds corresponding to NS5-branes 
with centers distributed at the circumference of a circle with
radius $r_0$ \cite{sfe1}. 
In the rest of this section we review the 
relevant to this paper results of \cite{sfe1} where the reader is referred for
further details.
The centers of the NS5-branes are located at 
\be
{\bf x}_i = r_0 (0,0,\cos\phi_i ,\sin\phi_i)\ ,\qq \phi_i=2\pi{i\ov N}\ ,\quad
i=0,1,2,\dots ,N-1\ .
\ee
In our presentation we will strictly restrict to the decoupling, 
near horizon limit, where the unit in \eqn{hkhhh} become unimportant.
We also keep finite 
the energies of strings with their 
ends attached between different centers.
Then the harmonic function is computed to be
\be
H=  N \L_N {1\ov \sqrt{(r^2+r_0^2)^2-4 r_0^2 \r^2}}\ ,
\label{hha}
\ee
where $\r^2=x_3^2+x_4^2$ and $r^2=x_1^2 + x_2^2 + \r^2 $  and
\be
\L_N= {\sinh N\chi \ov\cosh N\chi -\cos N \psi}\ ,
\ee
with the angular variable $\psi$ defined as $(x_3,x_4)=\r(\cos\psi,\sin\psi)$
and the auxiliary variable $\chi$ being given by
\be
e^\chi = {r^2+r_0^2\ov 2 r_0 \r }+\sqrt{\left(r^2+\r_0^2\ov 2 r_0 \r\right)^2
-1}\ .
\ee
When $N\gg 1$ and $r/r_0-1\gg 1/N$ the branes are practically 
continuously distributed 
in the circumference of the circle. 
Since then $\L_N\simeq 1$, the $Z_N$ symmetry 
becomes a continuous $SO(2)$ isometry. In the rest we will restrict to the 
continuum limit. It turns out that this is enough for the construction of our
plane wave solutions in the following section.

In order to exhibit the $SO(2)\times SO(2)$ symmetry explicitly we change 
variables as
\be
\pmatrix{x_1 \cr x_2} = r_0 \sinh \r \cos\th \pmatrix{\cos\phi\cr \sin \phi}
\ ,\qq \pmatrix{x_3\cr x_4} = 
r_0 \cosh \r \sin\th \pmatrix{\cos\psi\cr \sin \psi}  \ ,
\ee
where the range of the various variables is
\be 
0\le r<\infty\ , \qq 0\le \th\le {\pi\ov 2}\ ,\qq
0\le \psi\le 2\pi\ , \qq 0\le \phi\le 2\pi\ 
\ee
and also we rescale for future convenience
the $y_a$'s and $t$ by a factor of $\sqrt{N}$.
Then the background \eqn{skh} takes the form
\ba
&& {1\ov N}ds^2_{10} = \sum_{a=1}^5 dy_a^2  -dt^2 
+   d\r^2 +  d\th^2 +{1\ov 1 + \tanh^2{\r} \tan^2 \th }
\bl( \tan^2\th\ d\psi^2 + \tanh^2\r\ d\phi^2 \br)\ ,
 \nonumber\\
&&{1\ov N} B_{\phi\psi}= {1 \ov 1+ \tanh^2{\r} \tan^2{\th}}\ ,\quad
\label{backk}\\
&& e^{-2 \Phi} =e^{-2 \Phi_0} 
\bl(\cos^2{\th} \cosh^2 {\r} + \sin^2{\th} \sinh^2 {\r}
\br)\ .
\nonumber
\ea
Performing a T-duality transformation with respect to the Killing vector
${\del\ov \del\phi}$ we can relate this background to the one corresponding 
to the $SL(2,R)_{-N}/U(1)\times SU(2)_{N}/U(1)\times \IR^{1,5}$ exact theory. 

The metric in \eqn{backk} is singular at the location of the 
branes at the ring with coordinates $\rho=0$ and $\th=\pi/2$. The reason for 
this is that near the branes the continuous approximation breaks down and one
has to use the full solution \eqn{skh} with harmonic function given by
\eqn{hha}. Then one indeed verifies that the metric is non-singular 
\cite{sfe1}.

\subsection{The non-extremal NS5-brane rotating solution}

The background \eqn{backk} is the extremal supersymmetric limit of the 
most general non-extremal NS5-brane rotating solution constructed 
in \cite{sfe2} and further analyzed in \cite{HarmarkObers}.
In the field-theory limit this solution has a metric 
\ba
{1\ov N} ds^2 & = & -(1-{\m^2\ov \D_0}) dt^2 + \sum_{a=1}^5 dy_a^2 + {d\r^2 \ov
\r^2 + a_1^2 a_2^2/\r^2 + a_1^2 + a_2^2 -\m^2} 
\nonumber\\
 && + d\th^2 + {1 \ov \D_0}\left((\r^2+a_1^2)\sin^2\th d\phi_1^2 + 
(\r^2+a_2^2)\cos^2\th d\phi_2^2 \right)
\label{rot1}\\
&& -{2\ov \D_0}\m dt (a_1\sin^2 \th d\phi_1 + a_2\cos^2\th d\phi_2)\ ,
\nonumber
\ea
an antisymmetric tensor 
\be {1\ov N}B =  {2\ov \D_0}\left(-(\r^2+a_1^2)\cos^2\th d\phi_1\wedge d\phi_2
+ \m a_2 \sin^2\th dt\wedge d\phi_1 
+ \m a_1 \cos^2\th \m dt\wedge d\phi_2\right)\ ,
\ee
and a dilaton
\be
e^{-2\Phi} =  \D_0\ ,
\label{rott}
\ee
where we have defined
\be
\D_0=\r^2 + a_1^2\cos^2\th + a_2^2\sin^2\th\ 
\ee
and we have denoted by $a_1$, $a_2$ the angular parameters and by $\m$ the 
non-extremal parameter.
In was shown in \cite{sfe2} that for $\m=0$ this background becomes the
multicenter continuous distribution background in \eqn{backk} with
$r_0^2=|a_1^2-a_2^2|$. Moreover, it can be obtained by an $O(3,3)$ duality 
transformation on the background corresponding to the 
$SU(2) \times SL(2,\IR)/U(1)\times R^5$ exact CFT. Thermodynamic
properties of this solution can be found in \cite{sfe2,HarmarkObers}.

\section{The Penrose limit}
\setcounter{equation}{0}

In this section we will construct plane waves by taking Penrose
limits on the NS5 backgrounds \eqn{backk} and \eqn{rot1}. 

In general,
plane waves arise as classical solutions to theories of gravity 
and share the attractive feature that they depart from 
the flat space solution in the most controllable possible manner. 
This is essentially due to the existence of a covariantly constant 
null Killing vector, which, as it turns out,
guarantees that curvature effects are kept to a minimum 
\cite{kill1}-\cite{kill2}.
Plane waves, are relatively simple solutions which is reflected in the
simplicity of the equations of motions that they obey.
For a background, with only NS-NS fields turned on, of the form 
\ba
&&ds^2 = 2 dudv + F_{ij}(u) x_i x_j du^2 + dx_i^2 \ ,
\nonumber\\
&& B=B_{ij}(u) dx^i \wedge dx^j \ ,\qq \Phi=\Phi(u) \ ,
\label{brii}
\ea
the equations for 1-loop conformal invariance follow if the 
condition 
\be
-{\rm Tr}( F) + {1\ov 4} {\rm Tr} (S^2) +2 {d^2 \phi\ov du^2}=0\ ,
\qq  S_{ij}={dB_{ij}\ov du} \ ,
\ee
is satisfied.
It can be readily checked that this is indeed the case 
for all plane waves we construct below.
The conditions for conformal invariance at 1-loop are 
sufficient to prove conformal invariance to all orders in perturbation theory
\cite{Amati,Horowitz}.

\subsection{Plane waves from continuously distributed NS5-branes} 

We start with the Penrose limit for the background \eqn{backk}. 
According to the general prescription Penrose had suggested, 
we are supposed to magnify the space-time around a null geodesic. 
There are three natural such geodesics 
dictated essentially by symmetry. The first geodesic goes through the center 
of the ring corresponding to $\th=0$ in our coordinate parametrization.
The second geodesic corresponds to choosing $\th={\pi/2}$, therefore
laying on the plane of the ring which hits perpendicularly.
Finally, the third geodesic corresponds to $\r=0$ and hence also 
lays on the plane of the ring.
We recall that the background \eqn{backk} has a 
singularity at $\r=0 $ and $\th={\pi/2}$, corresponding to a breakdown
of the continuous approximation of the NS5-brane distribution.
Hence, the first geodesic never gets near the 
singularity of the background and will give rise to a completely non-singular 
plane wave. In contrast, the second and third 
geodesics hit the singularity 
and give rise to singular plane wave backgrounds. 
Nevertheless, the origin of this singularity is well understood as arising from
the breakdown of the continuous approximation at the location of the branes.
It is remarkable that, in some sense, 
this is captured by the solution giving rise to a seemingly well define
string spectrum in the light-cone gauge, as we will see in section 5.

\subsubsection{Geodesic at $\th=0$}

We first consider the geodesic with $\th=0$ that goes through the center of
the ring and has constant values for the 
space-like world-volume directions $y_a$.
In this case we have the effective 3-dim metric 
\be
ds_3^2 = - dt^2 + d\r^2 + \tanh^2 \r d\phi^2 \ .
\ee
We search for solutions to geodesic equations with $\rho,t,\phi$ being
functions of the proper time $\tau$.
Then, energy and $U(1)$-charge conservation imply that 
\be
\dot  t= E \ ,\qq \tanh^2\r \dot \phi  = l \ .
\ee
Using these conservation laws and the nullness of the geodesic we find that 
\be
\dot \r^2 = E^2 - l^2 \coth^2 \r \ .
\label{geod}
\ee
A real solution of this obviously exists, provided that the dimensionless
ratio $J=l/E$ obeys the condition $|J|\leq 1$.

Having such a null geodesic we define a new variable $u$ to be essentially 
the proper time $\tau$. Specifically, we choose $u=E \tau$ 
and therefore the meaningful parameter will be $J$. 
The other two variables parametrizing the directions
normal to the null geodesic in the $(t,\r,\phi)$ space 
will be denoted by $v$ and $x$. 
The appropriate change of variables in terms of exact differentials 
is explicitly given by 
\ba
&& d\r =\sqrt{1-J^2 \coth^2 \r } du \ ,
\nonumber\\
&& dt = du - dv/N + J dx/\sqrt{N}\ ,
\label{chava}\\
&& d\phi = J \coth^2\r du + dx/\sqrt{N} \ .
\nonumber
\ea
We note that the above change of variables requires that 
\be
\r\ge \ha \ln\left({1+J\ov 1-J}\right)\ ,\qq |J|\leq 1\ .
\ee
Also let 
\be
\th = {z\ov \sqrt{N}} \ ,\qq  y_{a} \to y_{a}/\sqrt{N}\ 
\ee
and define $dz^2 + z^2 d\psi^2 = d\vec z^2 $, so that $z^2 =\vec z^2$.
The relation between the variables $\r$ and $u$ is
\be
\cosh \r(u)=  {1\ov \sqrt{1-J^2}} \cosh\sqrt{1-J^2} u\ .
\ee
Then the background in the limit $N\to \infty $ takes the form
\ba
&& ds^2 = 2 du dv  + \sum_{i=1}^5 dx_i^2 + d\vec z^2 + (1-J^2) 
\tanh^2 \sqrt{1-J^2} u  dx^2 - J^2 \vec z^2 du^2 \ .
\nonumber\\
&& B_{12} = 2 J u \ ,
\label{back2}\\
&& e^{-2 \Phi}  =e^{-2 \Phi_0} \cosh^2 \sqrt{1-J^2} u \ .
\nonumber
\ea
Going to the Brinkman coordinates\footnote{
If the metric has the block-diagonal form 
\be
ds^2 = 2 du dv + \sum_i g_i(u) dx_i^2 \ ,
\ee
we obtain a metric in the Brinkman form \eqn{brii} 
with $F_{ij}=F_i(u)\d_{ij}$ after performing the transformation 
\be
u\to u \ , \qq v\to v+{1\ov 4} \sum_i {\dot g_i\ov g_i}\ ,
\qq F_i={1\ov 4}{\dot g_i^2\ov 
g_i^2} + \ha {d\ov du} \left(\dot g_i\ov g_i\right)\ .
\ee 
}
we find 
\ba
&& ds^2 = 2 du dv  + \sum_{i=1}^5 dx_i^2 + dx^2 + d\vec z^2 -\left(
2 {1-J^2\ov \cosh^2 \sqrt{1-J^2} u} x^2 + J^2 \vec z^2\right) du^2 \ ,
\nonumber\\
&& B_{12} = 2 J u \ ,
\label{back21}\\
&& e^{-2 \Phi}  =e^{-2 \Phi_0} \cosh^2 \sqrt{1-J^2} u \ .
\nonumber
\ea 
The parameter $J$ takes values in the interval $[-1,1]$. 
At the end points of this interval we recognize two known cases.
For $J=0$ the non-trivial part of our background is the
three-dimensional plane wave obtained in \cite{sfetse,BS1} as the Penrose limit
of the background for the $SU(2,\IR)_N/U(1)\times U(1)_{-N}$ exact CFT.
At the endpoints with $|J|=1$ we obtain as a non-trivial part of our 
background the four-dimensional plane wave of \cite{NW} constructed as a WZW 
model based on the semi-simple group $H_4$. In 
our context, this background arises as the Penrose limit of the background 
for the $SU(2)_N \times U(1)_{-N}$ exact CFT \cite{sfe22}.
For general values of $J$ we have a non-trivial five-dimensional 
theory which, as we will show, it can be obtained as a
contraction of the $SU(2)_N \times SL(2,\IR)_{-N'}/U(1)$ exact CFT.

\subsubsection{Geodesic at $\rho=0$}

In this case as well the geodesic lays on the plane of the ring and 
can be considered in parallel to the previous one.
Skipping some of the details, we first perform the change of variables 
\ba
&& d\th =\sqrt{1-J^2 \cot^2 \th } du \ ,
\nonumber\\
&& dt = du - dv/N + J dx/\sqrt{N}\ ,
\label{chava2}\\
&& d\psi = J \cot^2\th du + dx/\sqrt{N} \ .
\nonumber
\ea
Then we let 
\be
\rho = {z\ov \sqrt{N}} \ ,\qq  y_{a} \to y_{a}/\sqrt{N}\ 
\ee
and define $dz^2 + z^2 d\phi^2 = d\vec z^2 $, so that $z^2 =\vec z^2$.
The explicit relation between the variables $u$ and $\th$ is
\be
\cos \th=  {1\ov \sqrt{1+J^2}} \cos\sqrt{1+J^2} u\ .
\ee
Then the background in the limit $N\to \infty $ takes the form
\ba
&& ds^2 = 2 du dv  + \sum_{i=1}^5 dx_i^2 + dx^2 + d\vec z^2 + (1+J^2) 
\tan^2 \sqrt{1+J^2} u  dx^2 - J^2 \vec z^2 du^2 \ .
\nonumber\\
&& B_{12} = 2 J u \ ,
\label{back222}\\
&& e^{-2 \Phi}  =e^{-2 \Phi_0} \cos^2 \sqrt{1+J^2} u \ .
\nonumber
\ea
In Brinkman coordinates it reads
\ba
&& ds^2 = 2 du dv  + \sum_{i=1}^5 dx_i^2 + d\vec z^2 +\left(
2 {1+J^2\ov \cos^2 \sqrt{1+J^2} u}  x^2 - J^2 \vec z^2\right) du^2 \ ,
\nonumber\\
&& B_{12} = 2 J u \ ,
\label{back22}\\
&& e^{-2 \Phi}  =e^{-2 \Phi_0} \cos^2 \sqrt{1+J^2} u \ .
\nonumber
\ea 
We also note that \eqn{back222} and \eqn{back22} 
can can be obtained from \eqn{back2} and \eqn{back21} 
by analytically continuing as
$J\to i J $, $u\to i u$, $v\to -i v$ and $x\to i x$. As we have mentioned,
the singularity at 
$\sqrt{1+J^2} u = \pi/2$ originates from the breakdown of the continuous 
approximation near the location of the NS5-branes.

The plane wave, corresponding to the remaining geodesic with $\th=\pi/2$
will not be presented in this subsection since it is a particular case of 
the plane waves constructed in subsection 3.2 below.

\subsection{PP-wave limit of the non-extremal rotating NS5-brane}

It turns out that,
if we start with the most general rotating NS5-brane solution in the
field theory limit \eqn{rot1}-\eqn{rott}, it is enough 
to consider the null geodesic corresponding to $\th=\pi/2$.
For instance, a the null geodesics with $\th=0$ and $\th=\pi/2$ are
related by the symmetry transformation of the background 
under $\th\to \pi/2-\th$,
followed by the interchange of $a_1$ with $a_2$ and $\phi_1$ with $\phi_2$.
The 3-dim effective metric is then 
\be
ds_3^2 = 
-G_{tt} dt^2 + G_{\r\r} d\r^2 + G_{\phi_1\phi_1} d\phi_1^2 
-2 G_{t\phi_1} dt d\phi_1\ ,
\ee
where the various functions are given by 
\ba
&& G_{tt} = 1-{\m^2\ov \r^2 + a_2^2}\ ,
\qq G_{\r\r}\inv = {\r^2 +a_1^2 a_2^2/\r^2 +a_1^2 +a_2^2  -\m^2}\ ,
\nonumber\\
&& 
G_{\phi_1\phi_1} = {\r^2+a_1^2\ov \r^2+a_2^2}\ ,
\qq G_{t\phi_1}= \m {a_1\ov \r^2 + a_2^2}\ .
\ea
The energy and $U(1)$-change conservation laws give 
\be
G_{tt} \dot t + G_{t\phi_1} \dot \phi_1 = 1\ ,\qq G_{\phi_1\phi_1} \dot \phi_1 
- G_{t\phi_1} \dot t = J\ ,
\ee
with solution
\ba
&& \dot t =  {G_{\phi_1\phi_1} - J G_{t\phi_1} \ov G_{t\phi_1}^2 
+ G_{tt} G_{\phi_1\phi_1} }
= {(\r^2+a_2^2)(\r^2+a_1^2-J \m a_1)\ov \r^4 +(a_1^2+a_2^2-\m^2)\r^2 
+a_1^2 a_2^2}
\nonumber\\
&& \dot \phi_1 =  {G_{t\phi_1} + J G_{tt} \ov G_{t\phi_1}^2 + G_{tt} 
G_{\phi_1\phi_1}}
= {(\r^2+a_2^2)(\m a_1+J (\r^2+a_2^2-\m^2))
\ov \r^4 +(a_1^2+a_2^2-\m^2)\r^2 +a_1^2 a_2^2}\ .
\label{sjh2}
\ea
In the above we understand that all of the expressions when are given in 
terms of general functions $G_{tt}$ etc. hold in general.
Then, the nullness of the geodesic leads to 
\be
\dot \r^2 = {G_{\phi_1\phi_1} -2 J G_{t\phi_1} -J^2 G_{tt}\ov
G_{\r\r} (G_{t\phi_1}^2 + G_{tt} G_{\phi_1\phi_1})} = 
\left(1+{\a_2^2\ov \r^2} \right)\left[(J\m-a_1)^2
-J^2 a_2^2+(1-J^2)\r^2\right]\ .
\label{sjh}
\ee
We then perform the change of variables
\ba
&& d\r =\dot \r du \ ,
\nonumber\\
&& dt = \dot t du - dv/N + J dx/\sqrt{N} \ ,
\label{chava3}\\
&& d\phi_1 =\dot \phi_1 du + dx/\sqrt{N} \ ,
\nonumber
\ea
where $\dot \r$, $\dot t$ and $\dot \phi_1$ denote 
the functions in \eqn{sjh} and \eqn{sjh2} and also let
\be
\th = {\pi\ov 2}-{z\ov \sqrt{N}} \ ,\qq x_i\to x_i/\sqrt{N}\ .
\ee
Then, in the limit $N\to \infty$ we obtain the background
\ba
&& ds^2= 2 dudv +\sum_{a=1}^5 dy_a^2 + d\vec z^2 + \left(1-J^2
+{(a_1-J \m)^2-a_2^2\ov \r^2+a_2^2}\right)
 dx^2 -J^2 \vec z^2 du^2\ ,
\nonumber\\
&& B_{12}= -2 J u\ ,
\label{jw}\\
&& e^{-2\Phi}= \r^2 + a_2^2\ ,
\nonumber
\ea
where for the antisymmetric tensor we have used the gauge freedom in its
definition.

Let us first note that if the relation between the various parameters
$(a_1-J\m)^2=a_2^2$ holds, then we have the case of the four-dimensional 
plane wave of \cite{NW} times $\IR^6$.
Also, if we shift $a_1\to a_1+J\m$ the effect of 
non-extremality disappears from the above background as well as from the 
differential eq. \eqn{sjh}. Solving the latter we obtain a 
metric of the form \eqn{brii} with diagonal $F_{ij}$. In all cases 
$F_{1}=F_2=-J^2$, whereas the function $F_x$ and the dilaton 
depend on the range of various parameters.
We have three different cases (we use the shifted $a_1$ below):

\no
\underline{$a_1^2>a_2^2$ and $J^2< 1$}: Then 
\be 
G_{xx}=(1-J^2) \coth^2\sqrt{1-J^2}u \ ,\qq e^{-2\Phi}=e^{-2 \Phi_0} 
\sinh^2\sqrt{1-J^2} u\ .
\ee

\no
\underline{$a_1^2<a_2^2$ and $J^2< 1$}: Then 
\be 
G_{xx}=(1-J^2) \tanh^2\sqrt{1-J^2}u \ ,\qq e^{-2\Phi}=e^{-2 \Phi_0} 
\cosh^2\sqrt{1-J^2} u\ .
\ee

\no
\underline{$a_1^2>a_2^2$ and $J^2> 1$}: Then 
\be 
G_{xx}=(J^2-1) \cot^2\sqrt{J^2-1}u \ ,\qq e^{-2\Phi}=e^{-2 \Phi_0} 
\sin^2\sqrt{J^2-1} u\ .
\ee
All cases are related by analytic continuations of the various parameters
and variables.

Passing to the Brinkman coordinates we obtain a metric of the form 
\be
 ds^2= 2 dudv +\sum_{a=1}^5 dy_a^2 + dx^2+ d\vec z^2 + (F_x x^2 -J^2 
\vec z^2) du^2\ ,
\ee
with
\ba
&& F_x= 2 {1-J^2\ov \sinh^2 \sqrt{1-J^2} u }\ ,
\nonumber\\
&& F_x= -2 {1-J^2\ov \cosh^2 \sqrt{1-J^2} u }\ ,
\\
&& F_x= 2 {J^2-1\ov \sin^2 \sqrt{J^2-1} u }\ ,
\nonumber
\ea
corresponding to the three cases above, respectively.
Note that the last two cases correspond to plane waves identical to
the ones we obtained in subsection 3.1 (up to a trivial redefinition of $u$
in the trigonometric case). In addition, it can be checked that the first case 
is identical to the plane wave corresponding to the geodesic with $\th=\pi/2$
in the background \eqn{backk}.

We started with the non-supersymmetric,
non-extremal rotating NS5-background \eqn{rot1}-\eqn{rott} and obtained 
after the Penrose limit was taken, 
the same plane waves as those we obtained from the 
supersymmetric extremal background \eqn{backk} in the Penrose limit.
The physical reason for this is related to the fact that global information
about a space-time is lost after the Penrose limit is taken,
since essentially we focus and 
expand the space-time seen by a particular null geodesic.
Hence, there is no notion of a horizon in the plane wave geometry and
of the Hawking temperature associated with it.
This is related to the apparent loss of holography in the Penrose limit
(for discussion and some work in these directions see, for instance,
\cite{MaroHu,kirpio}). We believe that holography is present, but encoded in 
a string theoretical way beyond the supergravity approximation.
It can be explicitly checked, using results of \cite{sfe2,HarmarkObers},
that the Hawking temperature for the metric \eqn{rot1} goes to zero in 
the Penrose limit. Hence, 
states in the theory whose mass degeneracy is 
broken due to thermal effects, acquire again the same mass, resulting 
into a restoration of supersymmetry.
Having an exact CFT description 
in our case must be very important in working out the details of such 
a scenario.

\section{The relation to exact CFT}
\setcounter{equation}{0}

In this section we relate the plane waves \eqn{back21} and \eqn{back22}
to backgrounds 
corresponding to exact CFTs through a Penrose limit.
We also construct the corresponding symmetry algebra, via the associated
with the Penrose limit, contraction.

We start with the backgrounds corresponding to the 
$SU(2)_N\times SL(2,\IR)_{N'}/U(1) \times \IR^5 $ exact CFT 
\ba
&&ds^2 = N(d\th^2 + d\phi^2 + d\psi^2 + 2 \cos\th d\phi d\psi) +
N^\prime(-dt^2 + \tanh^2 t d\om^2) + \sum_{i=1}^5 dx_i^2\ ,
\nonumber\\
&& B_{\phi\psi}=N \cos\th\ ,
\label{kjsa}\\
&& e^{-2\phi}= e^{-2\Phi_0} \cosh^2 t\ .
\nonumber
\ea
Consider the change of variables
\ba
&&\th = {v\ov 2 J N} + 2 J u \ ,\qq t = \sqrt{1-J^2} u \ ,
\nonumber\\
&&\phi = {1\ov 2 \sqrt{N}} (x_1+x_2)\ ,\qq
\psi = {1\ov 2 \sqrt{N}} (x_1-x_2)\ ,\qq \om=\sqrt{N/N'\ov N'+N}\ x\ 
\label{dg1}
\ea
and take the limit 
\be
N,N'\to \infty\ ,\qq J^2={N'\ov N+N'}={\rm finite}\ .
\label{dg2}
\ee
We obtain
\ba
&& ds^2 = 2 du dv  + \sum_{i=1}^5 dx_i^2 + dx^2 + \cos^2 J u dx_1^2 
+ \sin^2 J u dx_2^2 + (1-J^2) \tanh^2 \sqrt{1-J^2} u \ ,
\nonumber\\
&& B_{12} = \cos 2 J u \ ,
\label{ba21}\\
&& e^{-2 \Phi}  =e^{-2 \Phi_0} \cosh^2 \sqrt{1-J^2} u \ .
\nonumber
\ea 
Passing to the Brinkman coordinates we obtain \eqn{back21}.
We also note that, had we started with the background 
for $SU(2)_{N'}/U(1) \times SL(2,\IR)_{-N} \times \IR^5$ 
exact CFT 
\ba
&&ds^2 = N(-dt + d\phi^2 + d\psi^2 + 2 \cosh t d\phi d\psi) +
N^\prime(d\th^2 + \tan^2 \th d\om^2) + \sum_{i=1}^5 dx_i^2\ ,
\nonumber\\
&& B_{\phi\psi}=N \cosh t\ ,
\label{kjsa1}\\
&& e^{-2\phi}= e^{-2\Phi_0} \cos^2 \th\ ,
\nonumber
\ea
we would have obtained in a Penrose limit, similar to \eqn{dg1}, \eqn{dg2},
the plane wave \eqn{back22}.

In order to construct the symmetry algebra for our planes waves \eqn{back21} 
and \eqn{back22} we should start with the symmetry algebras of the original 
backgrounds before the Penrose limit was taken.
We do that explicitly for the plane wave in \eqn{back22} and start with the 
symmetry algebra corresponding to the background \eqn{kjsa1}.
The $\IR^5$ factor corresponds to five free currents and therefore will be 
ignored in our discussion.
For the WZW model factor $SL(2,\IR)_{-N}$ we have a
current algebra theory generated by the three currents 
$J_0$, $J_\pm$ of conformal dimension $\D=1$ which obey the Operator Product
Expansions (OPE's)
\ba
&& J_0(z)J_0(w)=-{N/2 \ov (z-w)^2}+ {\rm regular}\ ,
\nonumber\\
&& J_0(z)J_\pm(w)={\pm J_\pm(w)\ov z-w} + {\rm regular}\ ,
\\
&& J_+(z)J_-(w)= - {2 J_0(w)\ov z-w}+{N\ov (z-w)^2} + {\rm regular}\ .
\ea
The symmetry algebra corresponding to the $SU(2)_{N'}/U(1)$ factor is
the parafermionic algebra of \cite{ZF}. 
This algebra is generated 
by chiral operators $\psi_l$, $l=0,1,\dots N'\!-\! 1$, with the complex
conjugation acting as $\psi^\dagger = \psi_{N'-l}$. Their conformal dimension
is $\D_l=l(N'-l)/N'$ and hence it differs from the classical result 
$\D_l^{\rm clas}=l$ (for finite $l$). 
This will have very important consequences for the nature 
of the exact CFT that follow from the contraction in the $N,N'\to \infty$
limit. Here we are particularly
interested for the two parafermionic operators $\psi_1$ and its
complex conjugate $\psi_1^\dagger$, with the lowest conformal dimension
$\D_1=1-1/N'$. Their OPE's are
\ba
&& \psi_1(z)\psi_1(w)={\sqrt{2(1-1/N^\prime)}
\ov (z-w)^{2/N^\prime}}(\psi_2(w)+{\cal O}(z-w))\ ,
\nonumber\\
&& \psi_1(z)\psi^\dagger_1(w)={1\ov (z-w)^{2\D_1}}
\left(1+ \left(1+{2\ov N^\prime}\right) 
(z-w)^2 T_{\rm par.}(w) + {\cal O}(z-w)^2\right)\ ,
\ea
where $T_{\rm par.}$ is the Virasoro stress energy tensor of the 
parafermionic theory.

\subsection{The contraction}

The quantum OPE's that we have exhibited, 
have their classical counterparts in terms of 
Poisson brackets. These are obeyed by the classical versions of the currents
$J_0,J_\pm$ and of the parafermions $\psi_1$ and $\psi^\dagger_1$ which are
realized in terms of the spacetime variables $t,\phi,\psi,\th$ and $\om$. 
Hence, the
Penrose limit for the background 
\eqn{back22} will give rise to a limiting procedure
for the corresponding symmetry generating classical objects. This is a 
quite straightforward, but lengthly, procedure and, as it turns out,
it gives rise to a contraction of the Poisson bracket algebra that 
the symmetry generators obey.
At this more abstract level the result can be taken over 
and be applied to the quantum case, where OPE's replace the classical 
Poisson brackets.
For the special case of the plane wave obtained as a Penrose limit
of the background for $SU(2)_N/U(1) \times U(1)_{-N}$ exact CFT this procedure 
was carried out in full detail in \cite{BS1}.
Here we skip all details and only quote the end result dealing 
directly with the quantum case. We define new operators $\Phi,\Psi,P,P_\pm$
as 
\ba
&& (1-J^2)^{1/2} \Phi  = {i\ov 2 \sqrt{N'}} (\psi_1-\psi^\dagger_1) 
-{1\ov N'} {J\ov \sqrt{1-J^2}} J_0 \ ,
\nonumber\\
&& (1-J^2)^{-1/2} \Psi  = {i \sqrt{N'}\ov 2 } (\psi_1+\psi^\dagger_1) +
{J\ov \sqrt{1-J^2}} J_0 \ ,
\nonumber\\
&& P={\psi_1+\psi^\dagger_1\ov 2}\ ,
\label{chha}\\
&& P_\pm = \sqrt{N} J_\pm \ .
\nonumber
\ea
In the limit \eqn{dg2}, all fields are primaries 
of the total Virasoro stress tensor with conformal dimension equal to one,
except for $\Psi$ for which we find  
\be
T(z)\Psi(w) = {\Psi(w)-\ha (1-J^2) \Psi(w)\ov (z-w)^2} + {\del\Psi(w)
\ov z-w} + {\rm regular}\ .
\label{hd1}
\ee
Hence, $\Psi$ is not a primary field unless $J^2=1$. In addition,
we find that 
\be
\Psi(z)\Psi(w)={1\ov (z-w)^2} \ , \qq \Phi(z)\Phi(w) = {\rm regular}\ .
\label{hd2}
\ee
These are the defining relations of a Logarithmic CFT \cite{Gur}.
The essential reason that a logarithmic structure for the CFT
arose is that the original 
theory contains as basic chiral operators the lowest parafermions with 
conformal dimension $1-1/N'$ as well as 
the $SL(2,\IR)_{-N}$ currents with dimension exactly $1$ for all $N$.
Hence, although classically all operators have dimension $1$, 
quantum mechanically this dimension is protected only in the case of the 
$SL(2,\IR)_{-N} $ currents. 
The remnant of this, in the limit $N',N\to \infty$, 
is the fact that $\Psi$ is not a primary field.
This phenomenon did not occur in the similar
contraction of current algebra theories \cite{Olive,sfe222}
since the combinations of
currents that one forms in order to take the limit $N\to \infty$,
have conformal dimension $1$ for all $N$.

The algebra of the logarithmic partners $(\Phi,\Psi)$ in \eqn{hd2} is enhanced
by the presence of non-trivial OPE's corresponding to the other fields 
in the theory. We find
\ba
&& 
P_+(z)P_-(w)={1\ov (z-w)^2} + J{\Phi(w)\ov z-w} + {\rm regular}\ ,
\nonumber\\
&& \Psi(z)P_\pm(w)=\pm  J{P_\pm(w)\ov z-w}+ 
{\rm regular}\ ,
\nonumber\\
&& \Psi(z)P(w)=-(1-J^2)\ln(z-w) (P\Phi)(w)+ {\rm regular}\ ,
\label{oper}\\
&& P(w)P(w)={1/2\ov (z-w)^2} +{1-J^2\ov 2}\ln(z-w)\Phi^2(w)+ {\rm regular}\ ,
\nonumber\\
&& \Psi(z)\Psi(w)= (1-J^2) \left[{\ln(z-w)\ov (z-w)^2}+2 \ln(z-w)
(P^2)(w) \right] 
\nonumber\\
&& \phantom{xxxxxxxxxx} + {(1-J^2)^2\ov 2} \ln^2(z-w) \Phi^2(w)
+ {\rm regular}\ .
\nonumber
\ea
For $J=0$ the operators $P_\pm$ become abelian currents and decouple from 
$\Phi$, $\Psi$ and $P$. The later obey the three-dimensional Logarithmic 
CFT found in 
\cite{BS1}. For $|J|=1$ the operator $P$ becomes an abelian current and 
decouples from $\Phi$, $\Psi$ and $P_\pm$ which, then, obey the current algebra
based on the non-semisimple group $H_4$ \cite{NW}. 
In this case we also note that $\Psi$ is a primary field. 
This limiting behaviour is in agreement with the
behaviour we have already noted for the background \eqn{back21}.

A quite straightforward generalization of our results to higher dimensional 
cases follows by replacing the coset factor $SU(2)_{N'}/U(1)$ with any compact
coset CFT. For the particular case of $SO(D+1)_{N'}/SO(D)$, this is immediate
using the results of the second ref. in \cite{BS1}.

\subsection{Free fields}

In order to reproduce the operator algebra \eqn{oper} using free fields,  
we utilize the free field realizations of the
coset model $SU(2)_{N'}/U(1)$ and of the $SL(2,\IR)_{-N}$ current algebra 
at finite $N,N'$ which are known \cite{parfree}. 
We start with the former model and introduce two real free bosons, 
\be
\langle \phi_i(z) \phi_j(w)\rangle = -\delta_{ij} \ln(z-w)\ ,\qq
i, j = 1, 2\ .
\ee
One may represent the elementary parafermion currents as\footnote{All 
expressions appearing in the rest of this paper to involve products of free
bosons and their derivatives, are understood as being properly normal ordered.}
\ba
\psi_1 &=& {1 \over \sqrt{2}}\left( -\sqrt{1 + {2/N'}}\ \partial \phi_1 
+ i \partial \phi_2 \right) e^{+\sqrt{2/N'}\ \phi_2}\ , 
\nonumber\\
\psi_1^\dagger &=& {1 \over \sqrt{2}} 
\left( \sqrt{1 + {2/N'}}\ \partial \phi_1 
+ i \partial \phi_2 \right) e^{-\sqrt{2/N'}\ \phi_2}\ .
\label{free1}
\ea
For the free field realization of the current algebra for $SL(2,\IR)_{-N}$
we need a time-like free boson $\phi_0$ and two spacelike free bosons 
$\varphi_i$, $i=1,2$. They obey 
\ba
&& 
\langle \varphi_i(z) \varphi_j(w)\rangle = -\d_{ij}\ln(z-w)\ , \qq i,j=1,2\ ,
\nonumber\\
&& 
\langle \phi_0(z) \phi_0(w)\rangle = \ln(z-w)\ .
\ea
Then the currents are given by\footnote{These are of course nothing but the
free field realization of the non-compact parafermions \cite{lykken} dressed 
with the extra boson $\phi_0$.}
\ba
&& J_\pm = \sqrt{N\ov 2}
\left( \mp\sqrt{1 - {2/N}}\ \partial \varphi_1 
+ i \partial \varphi_2 \right) e^{\mp i\sqrt{2/N} (\phi_0- \varphi_2)}\ .
\nonumber\\
&&J_0=i\sqrt{N\ov 2} \del \phi_0\ .
\label{free2}
\ea
The stress-energy tensor of the entire five-dimensional theory is 
\ba
T(z) & = &  -\ha (\del\phi_1)^2-\ha (\del\phi_2)^2 
-{i\ov \sqrt{2(N'+2)}} \del^2 \phi_1 \ 
\nonumber\\
& & + \ha (\del\phi_0)^2  - \ha (\del\varphi_1)^2 
- \ha (\del\varphi_2)^2 + {1\ov \sqrt{2(N-2)}} \del^2\varphi_1\ 
\label{strq}
\ea
and corresponds to a central charge $c=3N'/(N'+2)-1+3N/(N-2)$ theory,
as it should be.
Let us now consider the scalar field redefinition 
\ba
&&(1-J^2)^{1/2}\phi_+ = {1\ov \sqrt{2N'}} (\phi_0+\phi_1)\ ,\quad
(1-J^2)^{-1/2}\phi_- = \sqrt{N'\ov 2} (\phi_0-\phi_1)\ ,
\nonumber\\
&& \quad \phi_2= \phi\ ,\qq 
\varphi_\pm ={1\ov \sqrt{2}}(\mp\varphi_1+i\varphi_2)\ .
\label{coommm}
\ea
Then, the new set of scalars obey
\be
\langle \phi_+(z) \phi_-(w)\rangle 
=\langle \varphi_+(z) \varphi_-(w)\rangle 
= -\langle \phi(z) \phi(w)\rangle = 
\ln(z-w)\ 
\ee
and have zero correlators otherwise.
In order to take the $N,N'\to \infty$ limit, the expansion 
\be
\psi_1= -{\sqrt{N'}\ov 2} \del\phi_+ 
+ {1\ov \sqrt{2}} (i \del\phi-\phi \del \phi_+) +{1\ov \sqrt{N'}}\left(
{1\ov 2}\del\phi_- + i \phi\del\phi -\ha (\phi^2 + 1) \del\phi_+\right) 
+ {\cal O}\left(1\ov N'\right)\ ,
\ee
and its conjugate one for $\psi^\dagger_1$ are useful. After some algebra
we find the following free field representation for the basic operators of
our five-dimensional Logarithmic CFT 
\ba
&& \Phi=-i \del\phi_+ \ ,
\nonumber\\
&& P_\pm = \del\varphi_\pm e^{\mp i J \phi_+}\ ,
\nonumber\\
&& P={1\ov \sqrt{2}}\left(i \del\phi - \sqrt{1-J^2} \phi \del\phi_+\right)\ ,
\\
&& \Psi= i\del\phi_- -{i\ov 2}(1-J^2) (\phi^2+1) \del\phi_+ -\sqrt{1-J^2} \phi
\del\phi \ .
\nonumber
\ea
The Virasoro stress energy tensor expressed in terms of free fields becomes
\be
T(z) =  \del\phi_+\del\phi_- - \ha (\del\phi)^2 + \del\varphi_+ \del\varphi_-
-{i\ov 2 }\sqrt{1-J^2} \del^2\phi_+ \ 
\ee
and corresponds to a central change $c=5$ theory.
It is straightforward to check that these obey the OPE in \eqn{oper}.
We also note that for $J=0$ we obtain the free field representation 
of the three-dimensional Logarithmic CFT of \cite{BS1} and for $|J|=1$ that 
of the current algebra for the non-semisimple group $H_4$ found in \cite{KiKouLu}.

Using \eqn{chha}, the known correlation functions for the 
currents and of the parafermions and then taking the 
contraction limit, we may compute correlation functions 
for the basic operators of our theory $\Phi,\Psi,P$ and $P_\pm$. 
Alternatively, we may use the free field representation we have derived.
These computations will not be presented here, but the 
interested reader may get an idea of their form 
from the corresponding correlators in \cite{BS1}, where the parameter $J=0$.

\section{Strings in light-cone gauge}
\setcounter{equation}{0}

The purpose of this section is to explicitly solve for the 
string spectrum in the light-cone gauge. 
This is possible even though our plane wave backgrounds 
have $u$-dependent profiles. String theory on plane wave
backgrounds with only NS-NS fields turned on and time dependent profiles 
of the form $1/u^2$ have been solved in the light-cone gauge in 
\cite{Papadopoulos,Blau}. For extensive discussions of string on other
NS-NS backgrounds the reader is referred to \cite{Hikida,Sadri}.

\subsection{General formalism}

Let us first review the light-cone approach to string propagation in 
plane backgrounds (see, for instance, \cite{Amati,Horowitz})
suitably fitted for our purposes. 
We will consider bosonic backgrounds of the form
\eqn{brii} with diagonal $F_{ij}=F_i \d_{ij}$ and zero antisymmetric tensor.
This contains the case with $J=0$ for our backgrounds \eqn{back21} and
\eqn{back22}. The extension to cases with $J\neq 0$ is easily done and will
be briefly mentioned at the end of subsection 5.2.
We will consider the variable $u$ to be non-compact and let, 
as usual, $u=P \tau$. The light-cone action for the transverse coordinates is
\be
S={1\ov 4 \pi} \int d\tau d\s \left( \del_\tau X^i \del_\tau X^i -\del_\s X^{i}
\del_\s X^{ i} + P^2 F_i  X^i\right)\ .
\ee
The classical equations of motion are
\be
\d X^i :\qq -\del_\tau^2 X^i +\del_\s^2 X^{i} +P^2 F_i X^i =0\ 
\label{heg}
\ee
and 
\be
\Pi^i = {\d S\ov \d \dot X^i}={1\ov 2\pi} \del_\tau X^i \ ,
\ee 
denotes the conjugate to $X^i$ momenta. 
Using these the Hamiltonian takes the form
\be
H={1\ov 4\pi} \int_0^{2\pi} d\s 
\left( \del_\tau X^i \del_\tau X^i +\del_\s X^{i}
\del_\s X^{ i} - P^2 F_i  X^i\right)\ .
\label{haam}
\ee
We expand the string coordinates in Fourier modes as
\ba
X^i(\s,\tau) & = &  X^i_0(\tau) 
+ i \sum_{n\neq 0} {1\ov n}
e^{i n \s}\left(a_n^i X_n^i(\tau) - \tilde a_{-n}^i {X_n^i}^*(\tau)\right)
\nonumber\\
& = &  X^i_0(\tau) + i \sum_{n\neq 0} {1\ov n} X^i_n(\tau)
\left(a_n^i e^{i n \s} +  \tilde a_{n}^i e^{-i n \s}\right)
\nonumber\\
& = &  
X^i_0(\tau)+ i \sum_{n=1}^\infty {1\ov n} X^i_n(\tau)
(a^i_n e^{in\s} + \tilde a^i_n e^{-in\s}) 
\label{expp}\\
&&\phantom{xxxxxxxxx}
 - {1\ov n} X^i_{-n}(\tau) (a^i_{-n} e^{-in\s} + \tilde a^i_{-n} e^{in\s})\ ,
\nonumber
\ea 
where we have used ${a^*}^i_{n}=a^i_{-n}$ and ${X_n^i}^*=X^i_{-n}$, which 
follows from the reality condition for the $X^i$'s.
It will be convenient to identify a complex basis for the zero mode solution 
as
\be
X^i_0(\tau)=a^i \chi^i(\tau) + {a^i}^* {\chi^i}^*(\tau)\ .
\label{coobb}
\ee
Then using \eqn{heg} we find that the amplitudes obey 
the harmonic oscillator equations with $\tau$-dependent frequencies
\ba
&& \ddot X_n^i + (\om^i_n)^2 X_n^i = 0 \ ,
\qq \ddot \chi^i + (\om_0^i)^2\chi^i=0 \ ,
\nonumber\\
&& (\om^i_n)^2 = n^2 - P^2 F_i \ ,
\label{heg11}
\ea
where dots will denote ordinary derivatives with respect to $\tau$.
It is particularly useful to think of \eqn{heg11} as a Schr\"odinger
equation 
\be
-\ddot X_n^i +V_i X^i_n = n^2 X^i_n \ ,
\label{jdh}
\ee
where the potential is given by $V_i=P^2 F_i$.
For each $i$, the differential 
equation \eqn{heg11} has two independent solutions.
Their Wroskian will be normalized as 
\ba
&& \dot X^i_n  {X_n^i}^* - X^i_n {\dot X}_n^{i*} = i n\ ,\qq n\neq 0\ ,
\nonumber\\
&& \chi^i \dot \chi^{i*} - \dot\chi^i  \chi^{i*} = i \ .
\label{wwrr}
\ea
In a canonical quantization, one promotes 
$X^i, \Pi^i$, $a_n^i,\tilde a_n^i$ and $a^i,a^{i*}$ to operators and starts 
from the basic commutators 
\ba
&& [X^i(\s,\tau),X^j(\s',\tau)]=[\P^i(\s,\tau),\P^j(\s',\tau)]=0\ ,
\nonumber\\
&& [X^i(\s,\tau),\P^j(\s',\tau)]=i \d^{ij} \d(\s\!-\!\s')\ .
\ea
These give rise to 
\be
[a^i_n,a_m^j]=n \d^{ij}\d_{n+m}\ , \qq [a^i,a^{j\dagger}]=\d^{ij}\ ,
\ee
provided that the normalization \eqn{wwrr} is chosen.
Replacing the expansion \eqn{expp} 
into \eqn{haam} we obtain for the Hamiltonian
\be
H=H_0 + H_{\rm string} \ ,
\ee
where $H_0$ is the part of the Hamiltonian corresponding to the zero mode
\ba
 H_0 & =&   \ha \left(P_0^i P_0^i + (\om_0^i)^2 X_0^i X_0^i \right)
\nonumber\\
&= & 
\ha \left[\dot \chi^i\dot\chi^i+(\om^i_0)^2 \chi^i\chi^i\right] a^2
+\ha \left[\dot\chi^{i*}\dot\chi^{i*}+(\om^i_0)^2 \chi^{i*}\chi^{i*}\right]
(a^\dagger)^2
\label{zmh}\\
&& + \left[\dot\chi^{i}\dot\chi^{i*}+(\om^i_0)^2 \chi^{i}\chi^{i*}\right]
(a^{i\dagger} a^i+\ha) \ .
\nonumber
\ea
The relation to position and momentum operators is
\ba
&& X_0^i = a^i(\tau) \chi^i + {a^i(\tau)}^\dagger {\chi^i}^*\ ,
\nonumber\\
&& P_0^i = a^i(\tau) \dot\chi^i + {a^i(\tau)}^\dagger 
\dot\chi^{i*}\ .
\label{coobb1}
\ea
Since in the Schr\"odinger picture $X_0^i$ and $P_0^i$ are 
time-independent operators we have that 
\be 
{da^i\ov d\tau}={\del a^i\ov \del\tau}+i[H_0,a^i]=0\ ,
\ee
which can be proved by using \eqn{coobb1} and the equation of motion 
for $\chi^i$.
We have denoted by $H_{\rm string}$ the part of the Hamiltonian 
containing the string oscillation modes 
\ba 
&& H_{\rm string}  =  
\ha \sum_{n=1}^\infty {1\ov n^2}\Bigl[\left(\dot X^i_n \dot X^i_{-n}
+ (\om^i_n)^2 X^i_n X^i_{-n}\right)
(a^i_n a^i_{-n}  + \tilde a^i_n \tilde a^i_{-n})
\nonumber\\
&& \phantom{}
- \left(\dot X^i_n \dot X^i_{n} + (\om^i_n)^2 X^i_n X^i_{n}\right) a^i_n 
\tilde a^i_n 
- \left(\dot X^i_{-n} \dot X^i_{-n} + (\om^i_n)^2 X^i_{-n} X^i_{-n}\right) 
a^i_{-n} \tilde a^i_{-n}\Bigr]\ .
\label{haap}
\ea 

In the rest of this subsection we suppress for simplicity the index $i$.

It is convenient to write $\chi=re^{i \phi}$ and 
$X_n=r_n e^{i\phi_n}$ and then consider the equations for the 
amplitudes $r$, $r_n$ and the phases $\phi$ and $\phi_n$. 
Using \eqn{heg11} and \eqn{wwrr} we find that 
\be
\ddot r+ \om_0^2 r -{1\ov 4r^3}=0 \ ,\qq \dot \phi = -{1\ov 2 r^2} \ ,
\label{amp}
\ee
as well as 
\be
\ddot r_n+ \om_n^2 r_n -{n^2\ov 4r_n^3}=0 \ ,\qq \dot \phi_n = {n\ov 2 r^2} \ .
\label{amps}
\ee

Let us also note that the stringy-part of the Hamiltonian
is not diagonal, but it can be diagonalized by means of the transformation 
\be
\pmatrix{A_n\cr \tilde A^\dagger_n}=\pmatrix{f_n & f_{-n}\cr 
-f^*_{-n}& -f^*_n}\pmatrix{ a_n\cr \tilde a_{-n}}\ 
\ee
and its conjugation, where
\be
f_n={1\ov n \sqrt{2 \om_n}} (\dot X_n + i \om_n X_n) 
= {1\ov n \sqrt{2 \om_n}} e^{i\phi_n}
\left[\dot r_n + i \left(\om_n r_n+{n\ov 2 r_n^2}\right)\right]\ . 
\ee
Then we have the commutation rules 
\be
[A_n,A_m^\dagger]=\d_{n-m}\ ,\qq [\tilde A_n,\tilde A_m^\dagger]=\d_{n-m}
\ee
and zero otherwise. Then the expansion \eqn{expp} becomes 
\be
X(\s,\tau)= X_0(\tau) + i \sum_{n=1}^\infty {1\ov \sqrt{2 \om_n}}
\left[ e^{i n\s} (A_n-\tilde A_n^\dagger) 
+  e^{-i n\s} (A_n^\dagger - \tilde A_n)\right]\ .
\ee
In addition, \eqn{haap} assumes the diagonal form 
\be
H_{\rm string}= \ha \sum_{n=1}^\infty \om_n (A_n^\dagger A_n 
+ \tilde A_n^\dagger \tilde A_n)\ .
\ee 

The previous discussion was purely bosonic and in the supersymmetric case
should be supplemented by the inclusion of world-sheet fermions. In addition,
we have the level matching conditions arising from the expression for the 
light-cone variable $v$ in terms of the transverse coordinates $X^i$.
As we are interested in some rather generic features for string propagation 
in our backgrounds such discussions will be omitted in this paper.
We also note that in the case of constant frequency $\om_n^2=n^2$ we have, 
according to our normalizations, that $X_n = e^{i n \tau}/\sqrt{2}$ and we
recover familiar results for strings propagation in flat Minkowski 
space-time in the light-cone gauge.

\subsubsection{The particle-like spectrum}

Let's us discuss in some detail the particle spectrum corresponding
to the zero-mode Hamiltonian $H_0$.
The problem at hand is to determine a complete set of solutions to the 
time-dependent Schr\"odinger equation 
\be 
(i\del_\tau - H_0)\bar \Psi_n(x,\tau) =0 \ ,
\label{opppe}
\ee
where $H_0$ is given in \eqn{zmh}. This Hamiltonian 
is that for a quantum oscillator with time dependent 
frequency $\om_0(\tau)$.  It is possible to solve \eqn{opppe}
for any $\om_0(\tau)$.\footnote{For 
the description of the method see \cite{kimlee} and references therein.
We also note that the method has been used recently in the present context 
in \cite{Papadopoulos,Gimon}.}
Using the oscillator operators $a$ and $a^\dagger$ 
we can define, at a given $\tau$ (and $i$), 
a Fock space of states as usual
\be
a|0,\tau\rangle= 0 \ ,\qq 
|n,\tau\rangle = {1\ov \sqrt{n!}} (a^\dagger)^n|0,\tau\rangle\ .
\label{sttt}
\ee
In configuration space the eigenstates 
$\Psi_n(x,\tau)=\langle x|n,\tau\rangle$ can be constructed in terms of 
Hermite polynomials. 
Using the representation of $a$ and $a^\dagger$ as first
order differential operators we find that the normalized to one zero-mode
eigenstate satisfies
\be
a|0,\tau\rangle = 0 \quad \Rightarrow \quad 
(\chi^* \del_x - i \dot \chi^* x)\Psi_0=0
\quad \Rightarrow \quad \Psi_0(x,\tau)
=(\sqrt{2 \pi} |\chi|)^{-1/2} e^{{i\dot \chi^*\ov 2 \chi*} x^2}\ .
\ee
Then using \eqn{sttt} and the defining differential relation 
for Hermite Polynomials, 
we find that in configuration space\footnote{We correct below an apparent 
typo in eq. (A8) of \cite{kimlee}.}
\ba
\Psi_n(x,\tau) & =&  
(\sqrt{2 \pi} 2^n 2^n |\chi|)^{-1/2} \left(\chi\ov \chi^*\right)^{n/2}
e^{{i\dot \chi^*\ov 2 \chi*} x^2}
H_n\left(x\ov \sqrt{2}|\chi|\right) 
\nonumber\\
& = &  
(\sqrt{2 \pi} 2^n n! r)^{-1/2} e^{i n \phi} 
e^{\left(i{\dot r\ov 2 r}-{1\ov 4 r^2}\right)x^2}
H_n\left(x\ov \sqrt{2} r\right)\ ,
\ea
where in the last step we have used \eqn{amp}.
The advantage of this space is that we can write down immediately 
states $|\bar n,\tau \rangle $ that obey the operator equation 
\eqn{opppe}, simply as 
\be
\bar \Psi_n(x,\tau) = e^{i\th(\tau)} \Psi_n(x,\tau) \ ,
\label{bbeig}
\ee
where the phase $\th(\tau)$ is independent of the quantum number $n$ and
is determined by integrating the equation
\be
\dot \th= -\dot r^2 - \om_0^2 r^2 \ .
\ee
We note that in the case of constant frequency we have: 
$\chi=e^{-i \om_0 \tau} /\sqrt{2 \om_0}$ and 
$\th=-\ha \om_0\tau$. Then the total time dependence in \eqn{bbeig} 
is in a phase factor which reads $-\om_0(n+\ha)\tau $, as it should be.

\subsection{Application for our plane wave backgrounds}

We now apply the general formalism for our backgrounds \eqn{back21} 
and \eqn{back22}.
For the modes corresponding to the spatial brane directions $y_a$
we get the expected eigenfrequencies $\om_{a n}^2=n^2$. For the
eigenfrequencies corresponding to the transverse coordinates $z_i$, $i=1,2$
the general formalism we have developed is not adequate since we assumed 
zero antisymmetric tensor field. However, it is easy to see that 
we get a system of two coupled differential equations
\be
{d^2 z_{in}\ov d\tau^2}+(n^2 +P^2J^2) z_{i n}=2 i PJ n \e_{ij}
z_{j n}\ ,
\ee
where the term on the right hand side is due to the non-vanishing antisymmetric
tensor.
We easily see that the combinations $z_{\pm n}=z_{1n}\pm i z_{2n}$ 
diagonalize the system. The corresponding eigenfrequencies are 
\be
\om_{\pm n}^2 = (n \pm P J)^2 \ 
\ee
and coincide with those found in \cite{Forgacs} for the light-cone 
treatment of strings in the plane wave background of \cite{NW}
(for a recent discussion that takes into account the world-sheet fermions 
see \cite{Matlock}). 

In the rest of this subsection we set the parameter $J=0$ and 
concentrate on the three-dimensional plane wave, arising from the non-trivial
parts of \eqn{back21} and \eqn{back22}.

For the modes corresponding to the coordinate $x$ we get 
a Schr\"odinger equation as in \eqn{jdh} with potential 
that depends on the specific plane wave background.
In particular, in the case of the background \eqn{back21} we have 
\be
V= -2 {P^2 \ov \cosh^2 (P \tau)}\ .
\label{ppp1}
\ee
This belongs to a class of reflectionless potentials which can support a finite
number of bounds states. The reason for this is that this potential and 
the corresponding eigenvalue problem is related to the problem with
constant potential 
within the context of supersymmetric quantum mechanics \cite{susyqm}.
In our case there is exactly one bound state.
However, it is not admissible to the perturbative string spectrum since 
it corresponds to setting $n$ equal to the imaginary unit and that 
destroys the periodicity in the spatial string world-sheet variable $\s$.
The explicit solution for the states is (we follow the normalization 
of \eqn{wwrr})
\be
x_n(\tau) = {1\ov \sqrt{2}}
{i n - P \tanh a\tau\ov i n + P} e^{i n \tau}\ ,\qq n\neq 0\ ,\quad
-\infty <
\tau < +\infty\ ,
\ee
which clearly exhibits the reflectionless behaviour of the potential since the
wavefunctions at $\tau\to -\infty$ and at $\tau\to +\infty$ differ only by a 
phase factor.
Hence in our plane-wave background we dot not have string-mode creation.
For the string part of the Hamiltonian we get 
\be 
H_{\rm string}=\ha
\sum_{n=1}^\infty \G_n (a_n a_{-n} + \tilde a_n \tilde a_{-n})
+ \D_n a_n \tilde a_n + \D_n^* a_{-n} \tilde a_{-n}\ ,
\ee
where 
\ba
&& 
\G_n = 1+{P^2\ov n^2}{1\ov \cosh^2 P\tau} -{P^4\ov 2 n^2 (P^2+
n^2)} {1\ov \cosh^4 P\tau}\ ,
\nonumber\\
&&
\D_n= {1\ov 2} {P^3\ov n^2} {e^{2 i n \tau} \ov \cosh^4 P\tau}
{P \cosh 2 P\tau -i n \sinh 2 P \tau\ov (P+i n)^2} \ .
\ea
For $\tau\to \pm \infty$ this part of the
Hamiltonian becomes diagonal and takes the same form as in the case
of strings in flat space, i.e. $H_{\rm string}(\pm\infty)=
\ha \sum_{n=1}^\infty (a_n a_{-n} + \tilde a_n \tilde a_{-n})$. 

For the zero mode we have the, properly normalized, solution 
\be
\chi (\tau)= (2 \sqrt{2} P)^{-1/2} (P\tau \tanh P\tau -1) + 
i (\sqrt{2} P)^{-1/2} \tanh P\tau\ ,
\label{clll}
\ee
from which we deduce the expressions for the
amplitude $r(\tau)$ and the phase $\phi(\tau)$
\ba
&& r^2 = {1\ov \sqrt{2} P}\left(\ha(P \tau \tanh P\tau-1)^2 + \tanh^2 P\tau
\right)\ ,
\nonumber\\
&& 
\tan\phi = {\sqrt{2} \tanh P\tau\ov P\tau \tanh P\tau -1}\ . 
\label{ammmp}
\ea
This Hamiltonian has of course the form \eqn{zmh}  
where the coefficients of the quadratic operators are 
quite complicated to written down explicitly. We note that the constants of 
integration in \eqn{clll} have been chosen so that the Hamiltonian at 
$\tau = 0$ is diagonal and has the 
form $H_0(\tau=0)=\sqrt{2} P (a^\dagger a + \ha)$. 
With this choice the solution
at $\tau \to \pm \infty$ represents outgoing freely moving particle states.

Of particular interest is the amplitude to find the particle at a state
$\bar \Psi_n(x,\tau)$ having started at $\tau\to -\infty$, where the 
potential vanishes, as a plane wave,
i.e. $\bar \Psi_p^{\rm in}(x,-\infty)={1\ov \sqrt{2\pi}} e^{i p x}$,
where $p$ is the momentum.
The matrix element describing this is given by
\be
 \langle \bar\Psi_n(\tau)| \bar\Psi_p^{\rm in}(-\infty)\rangle
 =  (2\pi \sqrt{2 \pi} 2^n n! r)^{-1/2} e^{-i ( n \phi+\th)} 
\int_{-\infty}^\infty dx e^{i p x - (i{\dot r \ov 2 r}+{1\ov 4 r^2}) x^2 }
H_n\left(x\ov \sqrt{2} r\right) \ .
\ee
The integral is computed using the generating function for Hermite
polynomials. We find 
\be
\langle \bar\Psi_n(\tau)| \bar\Psi_p^{\rm in}(-\infty)\rangle=
\left(2\ov \pi\right)^{1/4}\sqrt{r\ov 2^n n!} e^{-i(n\phi+\th)}
e^{i {2 p^2 r^3\dot r \ov 4 r^2 \dot r^2 +1}}
\left(2 r \dot r + i\ov 2 r \dot r - i\right)^{n/2}
e^{-\ha \xi_p^2} H_n(\xi_p)\ ,
\ee 
where $\xi_p$ is a time-dependent function
\be
\xi_p(\tau)= {\sqrt{2} r p\ov \sqrt{4 r^2 \dot r^2 +1}}\ .
\ee
Therefore the corresponding amplitude is 
\be
\big |\langle \bar\Psi_n(\tau)| \bar\Psi_p^{\rm in}(-\infty)\rangle \big|^2
= \left(2\ov \pi\right)^{1/2}{r\ov 2^n n!} e^{-\xi_p^2} H_n^2(\xi_p)\ .
\ee
This expression is completely general and holds for all time-dependent 
frequencies that go to zero at $\tau\to -\infty$. In our case we may find the
explicit $\tau$-dependence of the solution by computing 
$\xi_p$ using \eqn{ammmp}. The result is not particularly
simple and will not be written down here.

We end this section by briefly discussing the case of strings propagating in 
the background \eqn{back22}, in the light-cone gauge, which has some distinct
features. 
In this case the modes corresponding to the coordinate $x$
obey a Schr\"odinger equation with potential
\be
V= 2 {P^2 \ov \sin^2 (P  \tau)}\ ,\qq
0 \leq P  \tau \leq {\pi}\ ,
\label{ppp2}
\ee
where we have shifted $P \tau$ by $\pi/2$. 
Similarly to \eqn{ppp1}, the Schr\"odinger problem with this potential 
is related, via supersymmetric quantum mechanics,
to the same problem with constant potential in the same 
finite interval for $\tau$. It turns out that the wavefunctions that vanish 
at $\tau=0$ and $\tau =\pi/P$ are 
\be
x_{n,m}=
(m+2) P \cos\left((m+2) P\tau \right) - P \cot P\tau 
\sin\left( (m+2) P\tau\right) ,\qq m=0,1,\dots \ ,
\ee
provided that $n^2= (m+2)^2 P^2$. Hence the light-cone momentum $P$
is quantized in order 
for a non-trivial solution to exist and in particular it should be a 
rational number. Given such a number, for each value $m$ there is only one 
string mode $n$, given by the above relation, that can be excited. We also note
the absence of the zero mode corresponding to $n=0$.
The phenomenon of the quantization of the light-cone parameter $P$ in the 
present context was found before in plane waves constructed 
from continuously distributed D3-branes \cite{BrS}. 
Clearly, restricting to the finite interval $\tau\in (0,\pi/P)$
is related to the fact the singularities of the metric in \eqn{back22}, where
the light-cone gauge breaks down. We expect that a covariant 
quantization of the string will shed light on this issue.

\section{Concluding remarks}

We constructed plane wave backgrounds 
corresponding to Penrose limits of NS5-branes.
The latter have a transverse space 
symmetry group $SO(2)\times Z_N\in  SO(4)$ 
and give rise to time-dependent profiles for the plane wave solutions.
We identify the corresponding exact theory 
as the five-dimensional Logarithmic CFT
arising from the contraction of the 
$SU(2)_{N'}/U(1)\times SL(2,R)_{-N}$ exact CFT, times $\IR^5$. 
We constructed a free field representation for
this theory and studied explicitly string propagation and spectra 
in the light-cone gauge. In view of the delicate issues concerning the 
validity of a uniform choice for the light cone gauge and states with 
light-cone momentum $P=0$, it will be desirable to also 
develop the covariant approach and construct string scattering amplitudes. 
Work towards this direction
has been undertaken for the plane wave of \cite{NW}, corresponding to $|J|=1$ 
and a current algebra in our class of models, 
in \cite{KiKouLu} and more recently in \cite{appol}.
It will be desirable to develop a similar approach at least 
for the case with $J=0$, corresponding to a purely Logarithmic CFT.

One of our motivations in considering the Penrose limits of brane 
distributions was to extend the considerations of \cite{BMN} for 
the sector of $\cN=4$ SYM with high spin and dimension operators, to cases 
away from conformality as the starting point. 
In particular, one way to break the conformal symmetry is
by means of the six scalars in the theory 
acquiring vacuum expectation values (vev's). 
We expect that in such cases the BMN 
operators are the same, but they have to act on a different vacuum.
On the supergravity side the centers of the D3-branes are
distributed in the six-dimensional transverse to the branes space,
according to the the distribution of vev's on the gauge theory side,
resulting into 
a deformation of the $AdS_5 \times S^5$ space. A particular such case was 
considered in \cite{BrS}, where plane wave solutions 
were obtained via Penrose limits on the solution representing
D3-branes uniformly distributed on a disc. 
In this case it was possible to obtain, in the limit of very large vev's, 
the perturbative string spectra in the light-cone gauge. 
However, there were problems related to the singularity
arising due to the breakdown of the continuous approximation,
similar to the case of the background \eqn{back22} we have discussed in the
present paper.
However, if we start with D3-branes distributed on a circle of 
radius $r_0$, instead of a disc, it is possible to obtain a plane wave which 
is completely non-singular by choosing, as in the case of \eqn{backk}, a
geodesic that goes through the center of the ring. The result is quite
simple and here we just mention the case with zero angular
parameter. This is purely geometrical since, as it turns out,
the metric is given by 
\be
ds_{10}^2 = 2 du dv +  \sum_{i=1}^{8} dx_i^2 + 
\left(3 (x_1^2 + x_2^2 -x_3^2  - x_4^2-x_5^2 ) + x_6^2 +x_7^2 +x_8^2 \right)
{du^2 \ov (1+u^2)^2}\ ,
\ee 
and the self-dual five-form is zero! 
This solution may well serve as a starting point for investigating the Coulomb 
branch of $\cN=4$ SYM in a particular sector 
but beyound the supergravity limit.

\newpage

\section*{Acknowledgements}

I would like to acknowledge the financial support provided through the European
Community's Human Potential Programme under contracts
HPRN-CT-2000-00122 ``Superstring Theory'' and HPRN-CT-2000-00131
``Quantum Structure of Space-time'', the
support by the Greek State
Scholarships Foundation under the contract IKYDA-2001/22 ``Quantum Fields
and Strings'',
as well as NATO support by a Collaborative Linkage Grant under the
contract PST.CLG.978785 ``Algebraic and Geometrical Aspects of Conformal
Field Theories and Superstrings''.

\end{document}